\documentclass[aps,pra,twocolumn,amsmath,nofootinbib]{revtex4-1}
\usepackage{booktabs}
\usepackage[T1]{fontenc}
\usepackage{threeparttable}
\usepackage{epsfig}
\usepackage[colorlinks,linkcolor=blue,anchorcolor=blue,
            citecolor=blue,urlcolor=blue,
            breaklinks=true]{hyperref}
\usepackage{graphics}
\usepackage{subfigure}
\usepackage{float}
\usepackage{epstopdf}
\usepackage{color}
\usepackage{dcolumn}
\usepackage{bm}
\usepackage{cases}
\usepackage{appendix}
\usepackage{lineno,hyperref}
\usepackage{ulem}

\newcommand{\tb}{\textbf}

\begin{document}
\author{Run Cheng$^{1,2}$}
\author{Qian-Yi Wang$^{1}$}
\author{Yong-Long Wang$^{2}$}
\author{Hong-Shi Zong$^{1,3,4}$}
\email{zonghs@nju.edu.cn}
\address{$^{1}$ Department of Physics, Nanjing University, Nanjing 210093, P. R. China}
\address{$^{2}$ School of Physics and Electronic Engineering, Linyi University, Linyi 276005, P. R. China}
\address{$^{3}$ Joint Center for Particle, Nuclear Physics and Cosmology, Nanjing 210093, P. R. China} \address{$^{4}$ Department of Physics, Anhui Normal University, Wuhu 241000, P. R. China}

\title{Finite-Size Effects with Boundary Conditions on
Bose-Einstein Condensation}

\begin{abstract}
We investigate the statistical distribution for ideal Bose gases with constant particle density in the 3D box of volume $V=L^{3}$. By changing linear size $L$ and imposing different boundary conditions on the system, we present a numerical analysis on the characteristic temperature and condensate fraction, and find that the smaller linear size is efficient to increase the characteristic temperature and condensate fraction. Moreover, there is a singularity under the antiperiodic boundary condition.

\end{abstract}
\maketitle

\section{Introduction}
The Bose-Einstein condensation (BEC) is a purely quantum-statistical phase transition characterized by the appearance of macroscopic population in ground state below the critical temperature $T_{c}$ and it plays an important role in condensed matter~\cite{Cataliotti2001,Inguscio2005,Roati2008a,Wu2011,Adhikari2019,Deuchert2019,Urvoy2019,Schneider2020,Georgescu2020}, optics~\cite{Klaers2010,Hansen2016}, atomic and molecular physics~\cite{Goldstein1997,Jochim2003,Zwierlein2003,zhang2020atomic} and among others. It is emphasized that the transition actually occurs at the thermodynamic limit, or when the discrete level structure was approximated by a continuous density of states~\cite{Anderson1995,Davis1995,Ketterle1996,Bradley1997,Brankov2000,Zwierlein2003,Kristensen2019}.

However,
the experimental observations of BEC on cold gases~\cite{Anderson1995,Davis1995,Bradley1997,Andrews1997,Myatt1997,Hall1998,Miesner1999,Matthews1999,Papp2008,McCarron2011} were performed in the finite volume that neither of the above approximations methods seems to be inherently justified~\cite{Grossmann1995,Ketterle1996}. Subsequently, some scholars wonder whether this will lead to deviations from the theoretical predictions and begin to pay attention to BEC in finite systems~\cite{Ziff1977,Grossmann1995,Ketterle1996,Franzosi2010}. The boundary conditions are of great importance to finite systems, such as periodic, Neumann, and Dirichlet conditions~\cite{London1938,Krueger1968, Greenspoon1974,Ziff1977,Grossmann1995,Holthaus2002,Brankov2020}.
The results imply that the shift of the condensation temperature depends only on the total number of particles~\cite{Grossmann1995,Ketterle1996}.

From the theoretical point of view, the only requirement is that the Hamiltonian of the finite system should be Hermitian, and the above boundary conditions are just some special cases.
These inspire us to explore the physics of BEC in finite volume systems by focusing on the finite-size behaviors and twisted boundary conditions~\cite{Byers1961}. For simplicity, in this paper, we model the finite system with 3D box of volume $V=L^{3}$ that consist of ideal Bose gases, keeping particle density fixed. And we characterize BEC phase transition with characteristic temperature and condensate fraction, then numerically calculate them under different linear sizes and boundary conditions.

This paper is organized as follows. In Section 2, the specific formulas for the statistical distribution of ideal Bose gases are directly given. In Section 3, through numerical calculations, finite-size effects on characteristic temperature and condensate fraction are investigated. In Section 4, the changes on characteristic temperature and condensate fraction are also obtained in the presence of twisted boundary conditions. Finally, in Section 5, conclusions and discussions are briefly given.

\section{The ideal Bose gases in the cubic box}\label{Section2}
In this section, the BEC statistics of ideal Bose gases with fixed particle density $n$, which is confined in the cubic box of volume $V=L^{3}$ are directly given. According to the Bose-Einstein distribution, the population $N(\epsilon_{i})$ of a state with energy $\epsilon_{i}$ is:
\begin{equation}\label{1}
  N(\epsilon_{i})=\frac{1}{e^{\beta(\epsilon_{i}-\mu)}-1}=\frac{ze^{-\beta\epsilon_{i}}}{1-ze^{-\beta\epsilon_{i}}},
\end{equation}
the particle density $n$ is:
\begin{small}
\begin{equation}\label{statistic 1}
n=\frac{1}{V}\sum_{i}\frac{1}{e^{\beta(\epsilon_{i}-\mu)}-1}=\frac{1}{V}\sum_{i}\frac{ze^{-\beta\epsilon_{i}}}{1-ze^{-\beta\epsilon_{i}}},
\end{equation}
\end{small}
where $\beta = 1/(k_{B}T)$ and $k_{B}$ denotes the Boltzmann's constant. And the fugacity $z$ related to chemical potential $\mu$ can be expressed by $z$ = exp$(\beta\mu)$. By splitting off the ground-state particle density $n_{0}$, the finite sum over the excited states $n_{\epsilon}$ is replaced by an integral:
\begin{equation}\label{statistic 2}
n_{\epsilon}=n-n_{0}=\frac{1}{V}\sum_{j=1}^{\infty} z^{j} \int_{\epsilon}^{\infty} D(\epsilon) \exp (-j \beta \epsilon) d \epsilon,
\end{equation}
where $n_{0} = \frac{ze^{-\beta\epsilon_{0}}}{1-ze^{-\beta\epsilon_{0}}}$. $D(\epsilon)$ represents the density of states, usually taken to be $\frac{2\pi V}{h^{3}}(2m)^{\frac{3}{2}}\epsilon^{\frac{1}{2}}$, $h$ and $m$ denote the Planck constant and the mass of boson, respectively. Assuming the ground state $\epsilon_{0}$ as zero, the result is:
\begin{equation}\label{statistic 3}
\frac{1}{V}\frac{z}{1-z}+\frac{1}{\lambda^{3}}g_{3/2}(z)=n,
\end{equation}
with $\lambda=\frac{h}{\left(2 \pi m k_{B} T\right)^{1 / 2}}$, and the Bose function $g_{n}(z)= \sum_{m=1}^{\infty}z^{m}/m^{n}$.
The physical meaning of the second term in Eq.~\eqref{statistic 3} implies that as the fugacity $z$ reaches its maximum $\exp(\beta\epsilon_{0})$, the number of the excited particles reaches the maximum, and all particles that exceeding this maximum must drop into the ground state. Namely, the critical temperature $T_{c}$ can now be found by setting $n_{0}=0$ and $z=1$ as:
\begin{equation}\label{statistic 4}
\frac{2\pi}{h^{3}}(2m)^{\frac{3}{2}}\int_{0}^{\infty}\frac{\epsilon^{\frac{1}{2}}d\epsilon}{e^{\frac{\epsilon}{k_{B}T_{c}}}-1}=n.
\end{equation}

We now consider the extension of an ordinary BEC, by confining ideal Bose gases into a finite cubic box with linear size $L$. For finite volume systems, it is always crossover rather than phase transition.
We thus define characteristic temperature~\cite{Brankov2020} $T_{c,L}$ of the finite system with linear size $L$ by:
\begin{equation}\label{statistic 5}
  \frac{1}{V}\sum_{i}\frac{1}{z_{0}^{-1}e^{\epsilon_{i}/k_{B}T_{c,L}}-1}=n, \quad (i=1,2,\cdots)
\end{equation}
with $z_{0}$=exp$(\epsilon_{0}/k_{B}T_{c,L})$. This temperature becomes critical temperature
only in the limit when system size goes to infinity.
Additionally, as for small systems, boundary conditions are of crucial importance as they can affect the symmetries of the system and consequently modify the fundamental properties, such as ground state energies and conserved quantities~\cite{Zawadzki2017}. The components of energy and momentum are quantized when particular physical boundary conditions are imposed on the finite system, and the discussions of BEC statistics should use summation instead of integral.

For simplicity, we consider particles confined in a 3D cubic box of dimensions $L_{x}$, $L_{y}$, $L_{z}$ $\in$[-$\frac{L}{2}$, $\frac{L}{2}$]. Theoretically, in order to ensure the momentum operator $\hat{p_{x}}$, $\hat{p_{y}}$, $\hat{p_{z}}$ in the systems are Hermitian,
the quantum probability density $\rho = |\psi|^{2}$ should be constant on the boundary of the cubic box. Namely,
\begin{equation}\label{2}
  \rho(x, y, z = -L/2) = \rho(x, y, z = L/2).
\end{equation}
Therefore, the arbitrary wave function should meet $\psi_{k}(\frac{L}{2})/\psi_{k}(-\frac{L}{2})=e^{i\alpha}$.
These twisted boundary conditions are parameterized by a twist angle $\alpha$ at each boundary, with range $0\leq\alpha\leq2\pi$. The use of twisted boundary conditions for the BEC in the cubic box, which is equivalent to realizing the BEC in the presence of a constant background magnetic potential coupled with bosons. The twist angles characterizing the twisted boundary condition are only obtained at the discrete values of the eigenstates energies, which corresponds to the magnetic flux quanta $N\phi$. The stronger the magnetic potential is, the larger is the twisted angle. Especially, the discrete twist angles are related to the linear size $L$, energy levels are shifted and finite-volume corrections can be generated~\cite{Sachrajda2005}. Experimentally, according to the superfluid behavior of a BEC~\cite{Koenenberg2015a,Koenenberg2016}, the achievement of BEC in the cubic of side length $L$ characterized with twisted boundary conditions may be realized in term of the non-classical response of the system to an infinitesimal boost or rotation. The boost can be provided by imposing extremely small velocity field such as slow rotation of the cubic box and the rotation can be equivalently produced by introducing disclinations and screw dislocations~\cite{Friedel2008Disclinations}.
Therefore,
the allowed components of particle's momenta inside the box are $p_{x,y,z} = \frac{2\pi}{L}N+\frac{\alpha}{L}$, where $N$ is an integer. And the corresponding energy eigenvalues of bosons in the cubic box are expressed as~\cite{Karbowski2000,Dalfovo2005}:
\begin{equation}\label{energy1}
\epsilon_{n_{1}n_{2}n_{3}}=\frac{\hbar^{2}[(2 n_{1}\pi+\alpha)^{2}+(2 n_{2}\pi+\alpha)^{2}+(2 n_{3}\pi+\alpha)^{2}]}{2mL^{2}},
\end{equation}
where the quantum states are characterized by the quantum numbers $(n_{1}, n_{2}, n_{3})$, $( n_{1}, n_{2}, n_{3}=0, \pm1, \pm2,\cdots )$. Consequently, we can absorb the effect of $\alpha$ into twist boundary conditions for the wavefunctions, to study different physical boundary conditions~\cite{Kirsten1998} on BEC in finite volume systems. Substituting Eq.~\eqref{energy1} into Eq.~\eqref{statistic 5}, with the volume $V = L^{3}$, that as:
\begin{equation}\label{number of particles}
  \frac{1}{L^{3}}\sum_{n_{1}} \sum_{n_{2}}\sum_{n_{3}}\frac{1}{z_{0}^{-1}{e^{\frac{\epsilon_{n_{1}n_{2}n_{3}}}{k_{B}T_{c,L}}}-1}}=n.
\end{equation}
Note that the most significant difference among different boundary conditions is the difference in the ranges Eq.~\eqref{number of particles} in which the quantum numbers $(n_{1}, n_{2}, n_{3})$ vary.
Particularly, the ground state energy which depends sensitively on $\alpha$ needs to be excluded during the calculation. For clarity, here we define $\alpha=k\pi$ $($k$\in$ $[0,2]$$)$.

(i)$k\in[0,1)$, the ground state $(0,0,0)$  should be excluded.

(ii)$k\in(1,2]$, the ground state $(-1,-1,-1)$  should be excluded.

(iii)$k=1$, the ground state $(-1,-1,-1)$, $(-1,-1,0)$, $(-1,0,-1)$, $(0,-1,-1)$, $(0,0,-1)$, $(0,-1,0)$, $(-1,0,0)$, $(0,0,0)$ should be excluded.

Obviously, the BEC statistics in the finite volume
system not only depends on the linear size but also on the boundary
conditions. And we will do some detailed calculations in following sections.

\section{Finite-size effects on Bose-Einstein condensation in the cubic box}\label{Section3}
In this section, performing the BEC statistics for the finite system of ideal Bose gases in the box traps of different linear sizes under periodic, antiperiodic and Dirichlet boundary conditions, we compare their statistics and conclude on the finite-size effects on them.

(a) Periodic boundary condition ($\alpha = 0$)

\begin{figure}[htbp]
\centering
\includegraphics[width=0.46\textwidth]{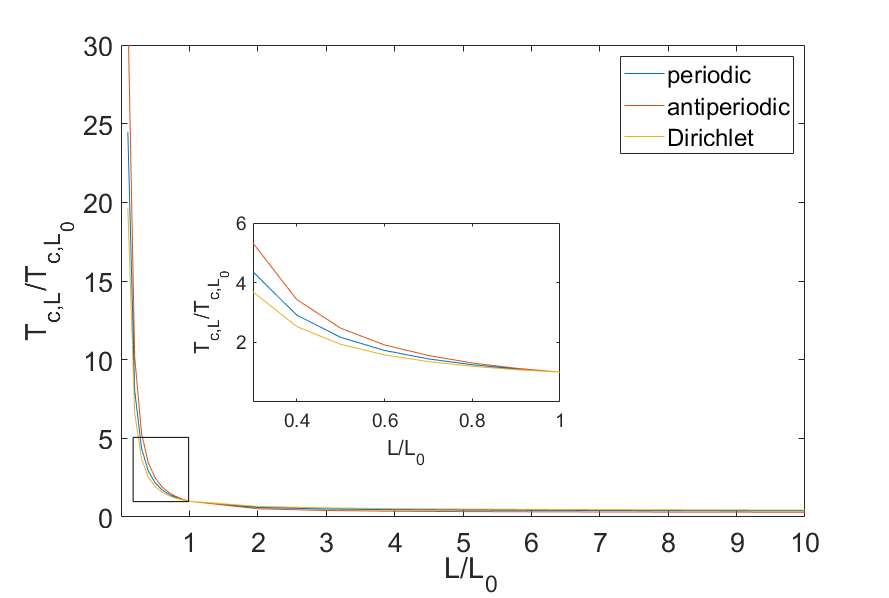}
\caption{\footnotesize (Color online) The characteristic temperature $T_{c,L}$ versus $L$ for an ideal Bose gases enclosed in the cubic box under the periodic, antiperiodic and Dirichlet boundary conditions. $L_{0}$ is the unit length, $T_{c,L_{0}}$ is the corresponding characteristic temperature.}\label{fig1}
\end{figure}

\begin{figure}[htbp]
\centering
\includegraphics[width=0.46\textwidth]{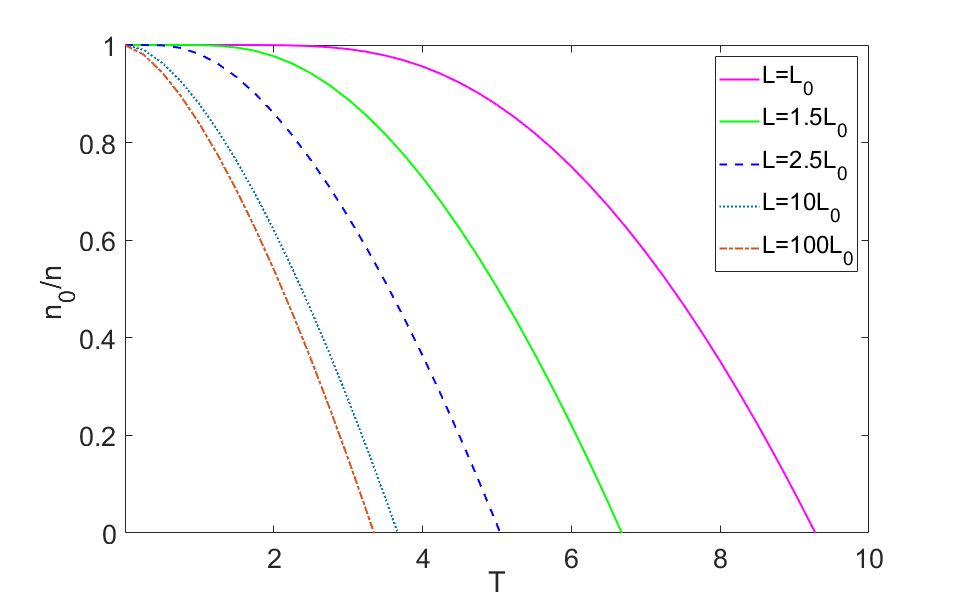}
\caption{\footnotesize (Color online) The ground state particle density $n_{0}/n$ versus $T$ in the cubic box, with $L=L_{0}, 1.5L_{0}, 2.5L_{0}, 10L_{0}, 100L_{0}$, under the periodic boundary condition. Here $\frac{\hbar^2}{mk_{B}}n^{\frac{2}{3}}$ is taken as an unit of $T$.}\label{fig4}
\end{figure}

According to our analyses, the discontinuous energy values of single-state can be given by:
\begin{equation}
\epsilon_{n_{1}n_{2}n_{3}}=\frac{2\hbar^{2}\pi^{2}(n_{1}^{2}+ n_{2}^{2}+ n_{3}^{2})}{mL^{2}}.
\end{equation}
The characteristic temperature can be determined by:
\begin{equation}\label{number of particles11}
  \frac{1}{L^{3}}\sum_{n_{1}} \sum_{n_{2}}\sum_{n_{3}}\frac{1}{e^{\frac{2\hbar^{2}\pi^{2} ( n_{1}^{2}+n_{2}^{2}+n_{3}^{2})}{mL^{2}k_{B}T_{c,L}}}-1}=n.
\end{equation}
The numerical results under the periodic boundary condition are shown in Figs. ~\ref{fig1}, ~\ref{fig4} and TABLE I.

(b) Antiperiodic boundary condition ($\alpha = \pi$)

\begin{figure}[htbp]
\centering
\includegraphics[width=0.46\textwidth]{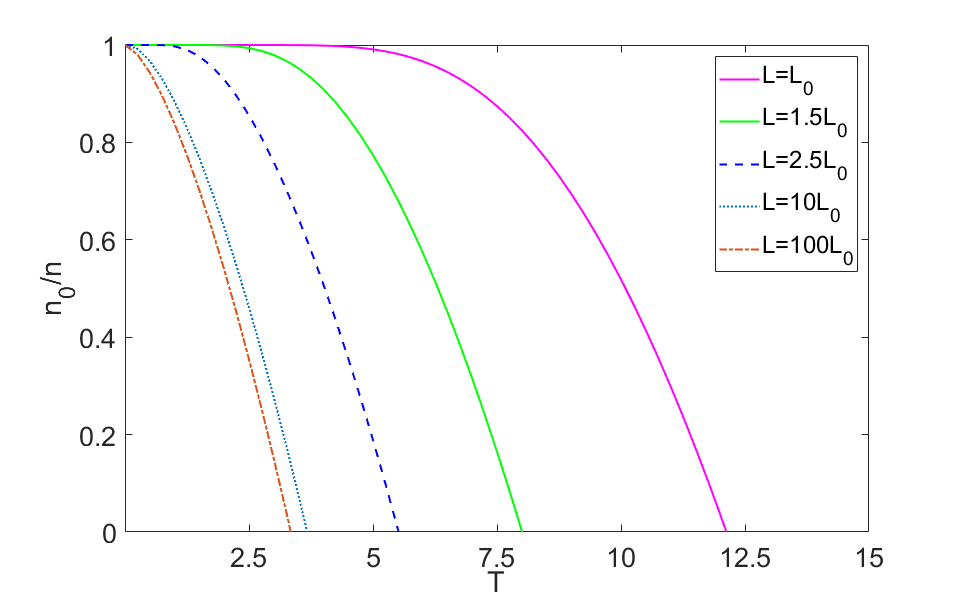}
\caption{\footnotesize (Color online) The ground state particle density $n_{0}/n$ versus $T$ in the cubic box, with $L=L_{0}, 1.5L_{0}, 2.5L_{0}, 10L_{0}, 100L_{0}$, under the counter-periodic boundary condition. Here $\frac{\hbar^2}{mk_{B}}n^{\frac{2}{3}}$ is taken as an unit of $T$.}\label{fig5}
\end{figure}

The quantum state energy is given by:
\begin{equation}
\epsilon_{n_{1}n_{2}n_{3}}=\frac{\hbar^{2}[(2 n_{1}\pi+\pi)^{2}+(2 n_{2}\pi+\pi)^{2}+(2 n_{3}\pi+\pi)^{2}]}{2mL^{2}}.
\end{equation}
As a result, the characteristic temperature can be calculated by:
\begin{equation}\label{number of particles21}
  \frac{1}{L^{3}}\sum_{n_{1}} \sum_{n_{2}}\sum_{n_{3}}\frac{1}{e^{\frac{\hbar^{2}\pi^{2}[( 2n_{1}+1)^{2}+(2n_{2}+1)^{2}+(2n_{3}+1)^{2}-3]}{2mL^{2}k_{B}T_{c,L}}}-1}=n.
\end{equation}
Similarly, the corresponding results are shown in Figs. ~\ref{fig1}, ~\ref{fig5} and Table I.

(c) Dirichlet condition

\begin{figure}[htbp]
\centering
\includegraphics[width=0.46\textwidth]{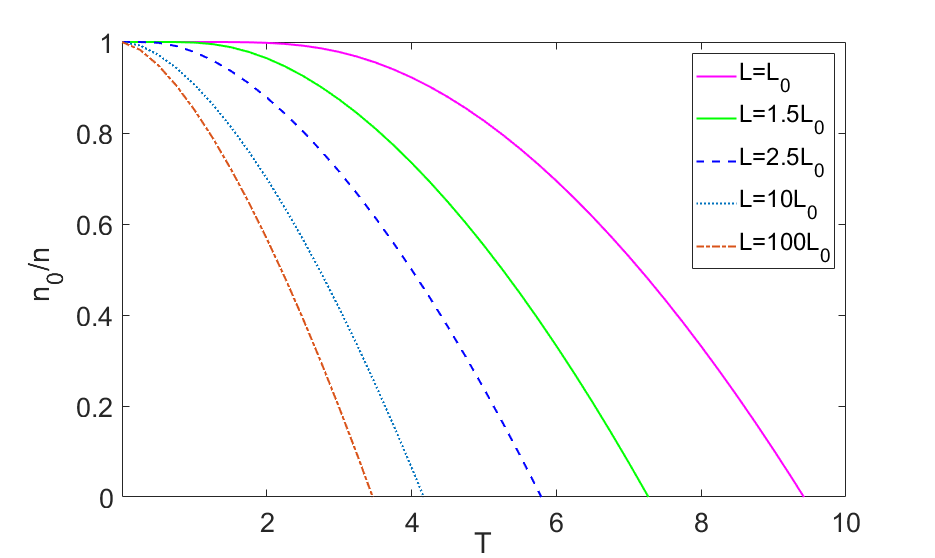}
\caption{\footnotesize (Color online) The ground state particle density $n_{0}/n$ versus $T$ in the cubic box, with $L=L_{0}, 1.5L_{0}, 2.5L_{0}, 10L_{0}, 100L_{0}$, under the Dirichlet boundary condition. Here $\frac{\hbar^2}{mk_{B}}n^{\frac{2}{3}}$ is taken as an unit of $T$.}\label{fig6}
\end{figure}

This can be described by requiring that the wave function of the particle is identically zero outside the box. Hence an impenetrable barrier can be interpreted as a boundary condition. The Dirichlet condition requires $\psi=0$ on each side of the cubic box, $\psi_{k}(\frac{L}{2})=\psi_{k}(-\frac{L}{2})=0$ $(k=x,y,z)$. In this case, the single-state energy:
\begin{equation}
\epsilon_{n_{1}n_{2}n_{3}}=\frac{\hbar^{2}\pi^{2}(n_{1}^{2}+ n_{2}^{2}+ n_{3}^{2})}{2mL^{2}}.
\end{equation}
The characteristic temperature can be derived by:
\begin{equation}\label{number of particles31}
 \frac{1}{L^{3}}\sum_{n_{1}=1}^{\infty} \sum_{n_{2}=1}^{\infty}\sum_{n_{3}=1}^{\infty}\frac{1}{e^\frac{\hbar^{2}\pi^{2}( n_{1}^{2}+n_{2}^{2}+n_{3}^{2}-3)}{2mL^{2}k_{B}T_{c,L}}-1}=n.
\end{equation}
Here noted that the ranges of the quantum numbers $(n_{1},n_{2},n_{3})$ under Dirichlet boundary condition are from 1 to $\infty$. Therefore, the nonvanishing ground state $(1,1,1)$, $\epsilon_{0}=\frac{3\hbar^{2}\pi^{2}}{2mL^{2}}$ should be excluded. The Figs. ~\ref{fig1}, ~\ref{fig6} and TABLE I depict the corresponding results.

\begin{table}[htbp]
 \tb{TABLE I. The characteristic temperature $T_{c,L}$ versus $L$ for an ideal Bose gases enclosed in the cubic box of $L=0.2L_{0}, 0.5L_{0}, 10L_{0}, 50L_{0}, 100L_{0}$,under the periodic, antiperiodic and Dirichlet boundary conditions.}
 \renewcommand\arraystretch{1.0}
  \renewcommand\tabcolsep{5.0pt}
\begin{center}
  \begin{tabular}{|c|c|c|c|c|c|c|}
  \hline
  {$L/L_{0}$} & {0.2} & {0.5} & {1} &{10} &{50} &{100}  \\
  \hline
     (a)$k_{B}T_{c,L}(\frac{\hbar^{2}}{m}{n}^{\frac{2}{3}})$ & 74.5 & 20.1 &9.279 &3.67 & 3.38 & 3.35\\
    \hline
    (b)$k_{B}T_{c,L}(\frac{\hbar^{2}}{m}{n}^{\frac{2}{3}})$ & 123.3 & 30 &12.13 &3.68 & 3.37 & 3.34\\
    \hline
     (c)$k_{B}T_{c,L}(\frac{\hbar^{2}}{m}{n}^{\frac{2}{3}})$  & 62.4 & 18.2 &9.42 &4.168 & 3.57 & 3.463\\
  \hline
\end{tabular}
\end{center}
\end{table}

By calculating numerically, the changes of $T_{c,L}$ with $L$, under periodic, antiperiodic and Dirichlet conditions are shown in Fig. ~\ref{fig1}.  It is easy to check that no matter which boundary condition is taken, at a fixed particle number density, $T_{c,L}$ considerably increases as $L$ decreases in the region $L/L_{0}<1$. Since $T_{c,L}$ describes an approach to condensed state from the non-condensed one, the smaller linear size allows BEC to occur at higher temperature. And when $L>10L_{0}$, the values of $T_{c,L}$ agree with the results obtained within the thermodynamic limit results~\cite{Holthaus2002}, as shown in TABLE I, where we have chosen $\frac{\hbar^{2}}{mk_{B}}n^{\frac{2}{3}}$ as the $T_{c,L}$ unit.
In order to illustrate this clearly, we numerically discuss the particle density for the ground state $n_{0}$ under the three boundary conditions sketched in Figs. ~\ref{fig4}, ~\ref{fig5} and ~\ref{fig6}, respectively. It is clear that as the linear size decreases the slope of the curves gradually becomes slower, condensate fraction $n_{0}/n$ all appears to increase. Namely, the smaller linear size is effective to increase the condensate fraction. Meanwhile, the intersection point of the curve and the horizontal axis $T_{c,L}$, being shifted to high value. The behavior further illustrates that $T_{c,L}$ describe the crossover behavior between ground and excited states, and $T_{c,L}$ becomes critical temperature $T_{c}$ when the volume of the system tend to infinity. These results due to the fact that the coupling between energy level and linear size as Eq.~\eqref{number of particles} can significantly alters the nature of BEC in the finite system. The smaller linear size can produce a stronger restrictions, and accordingly lower the mean energy, which can greatly facilitate the realization of BEC~\cite{Bagnato1987}, increasing $T_{c,L}$ and $n_{0}/n$.

Consequently, the finite-size effects play an important role in the small volume systems, allowing for $L-$dependence of the shift of $T_{c,L}$ and $n_{0}/n$. The smaller the linear size (the larger the confinement), the higher the $T_{c,L}$ and the greater the $n_{0}/n$.

\section{The Influences of boundary conditions on Bose-Einstein condensation in a cubic box}\label{Section4}

\begin{figure}[htbp]
\centering
\includegraphics[width=0.46\textwidth]{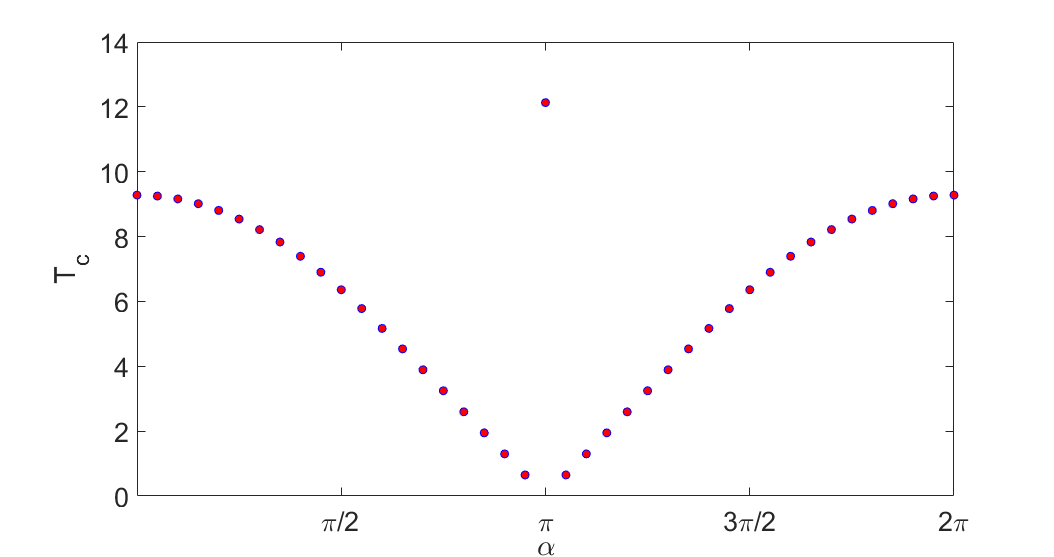}
\caption{\footnotesize (Color online) The characteristic temperature $T_{c,L}$ versus $\alpha$ with $L=L_{0}$ in the cubic box. Here $\frac{\hbar^2}{mk}n^{\frac{2}{3}}$ is taken as an unit of $T_{c,L}$.}\label{fig7}
\end{figure}

\begin{figure}[htbp]
\centering
\includegraphics[width=0.46\textwidth]{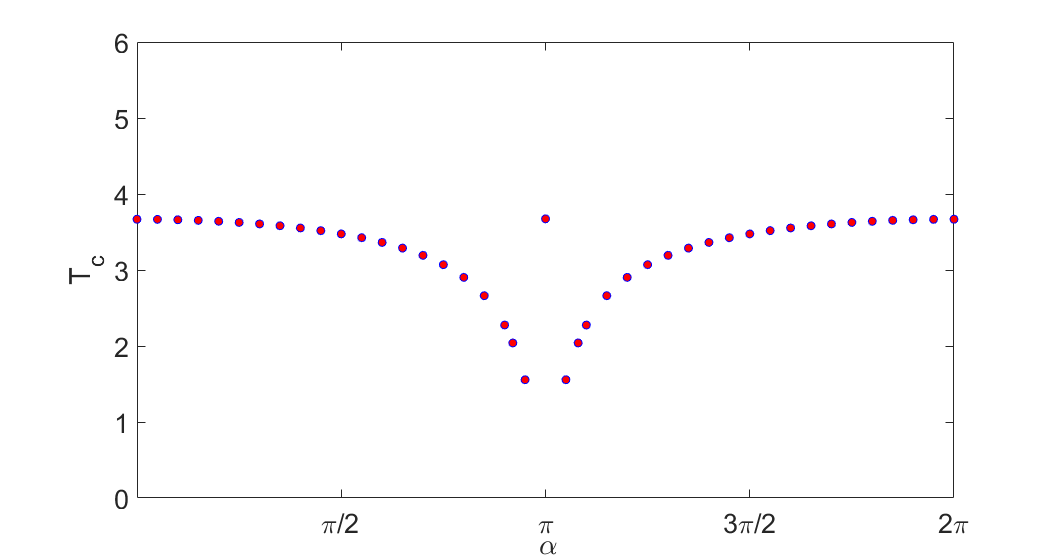}
\caption{\footnotesize (Color online) The characteristic temperature $T_{c,L}$ versus $\alpha$ with $L=10L_{0}$ in the cubic box. Here $\frac{\hbar^2}{mk}n^{\frac{2}{3}}$ is taken as an unit of $T_{c,L}$.}\label{fig8}
\end{figure}

\begin{figure}[htbp]
\centering
\includegraphics[width=0.46\textwidth]{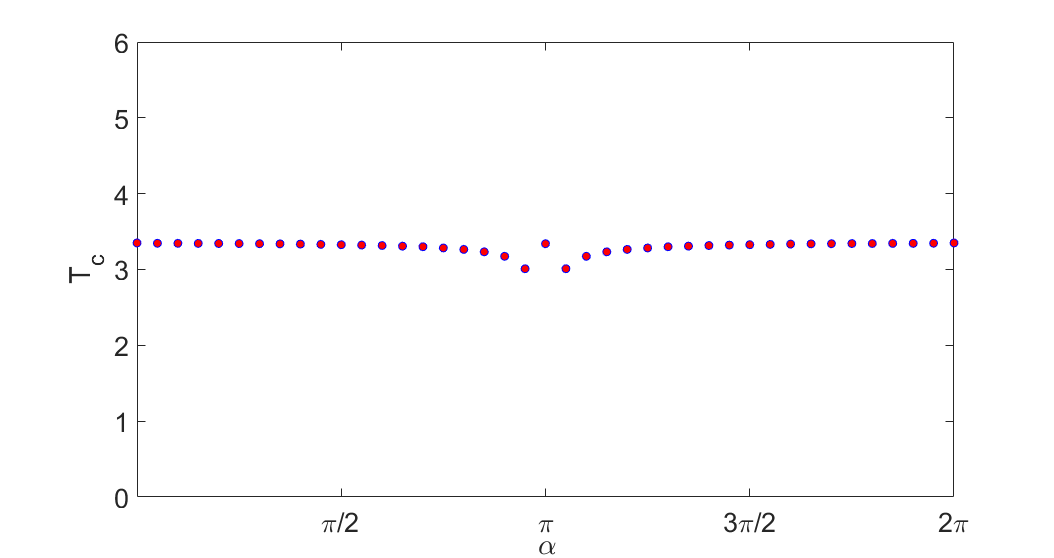}
\caption{\footnotesize (Color online)  The characteristic temperature $T_{c,L}$ versus $\alpha$ with $L=100L_{0}$ in the cubic box. Here $\frac{\hbar^2}{mk}n^{\frac{2}{3}}$ is taken as an unit of $T_{c,L}$.}\label{fig9}
\end{figure}

\begin{figure}[htbp]
\centering
\includegraphics[width=0.46\textwidth]{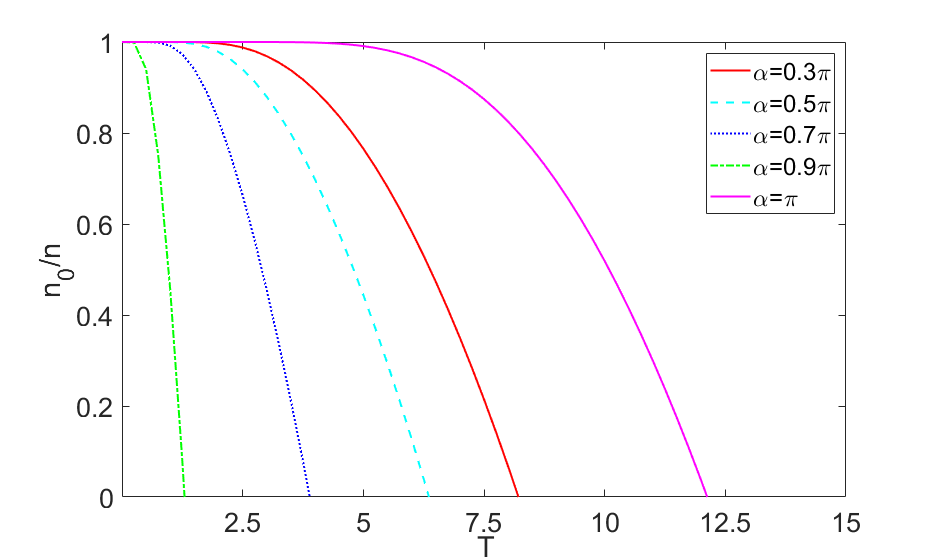}
\caption{\footnotesize (Color online) The ground state particle density $n_{0}/n$ versus $T$ with $L=L_{0}$, with $\alpha=0.3\pi, 0.5\pi, 0.7\pi, 0.9\pi, \pi$ in the cubic box. Here $\frac{\hbar^2}{mk}n^{\frac{2}{3}}$ is taken as an unit of $T$.}\label{fig10}
\end{figure}

\begin{figure}[htbp]
\centering
\includegraphics[width=0.46\textwidth]{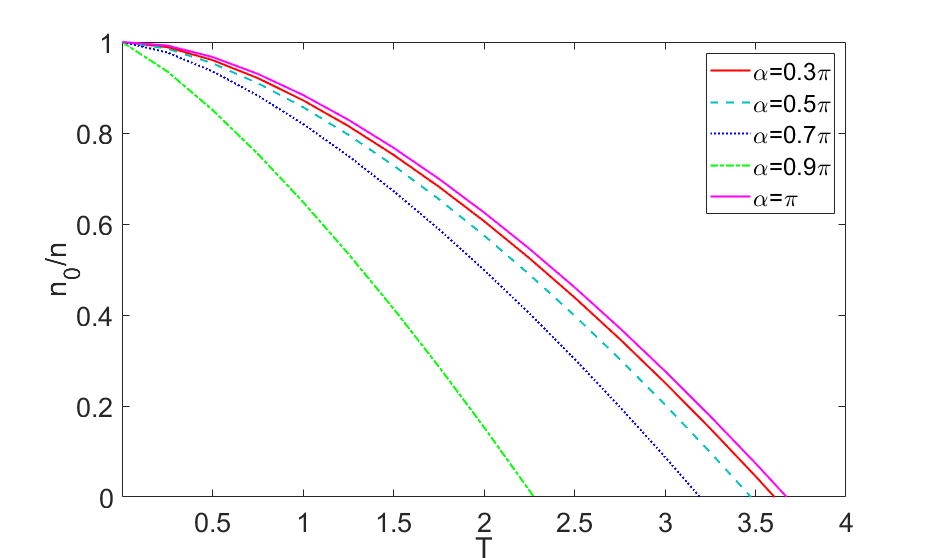}
\caption{\footnotesize (Color online) The ground state particle density $n_{0}/n$ versus $T$ with $L=10L_{0}$, with $\alpha=0.3\pi, 0.5\pi, 0.7\pi, 0.9\pi, \pi$ in the cubic box. Here $\frac{\hbar^2}{mk}n^{\frac{2}{3}}$ is taken as an unit of $T$.}\label{fig11}
\end{figure}

\begin{figure}[htbp]
\centering
\includegraphics[width=0.46\textwidth]{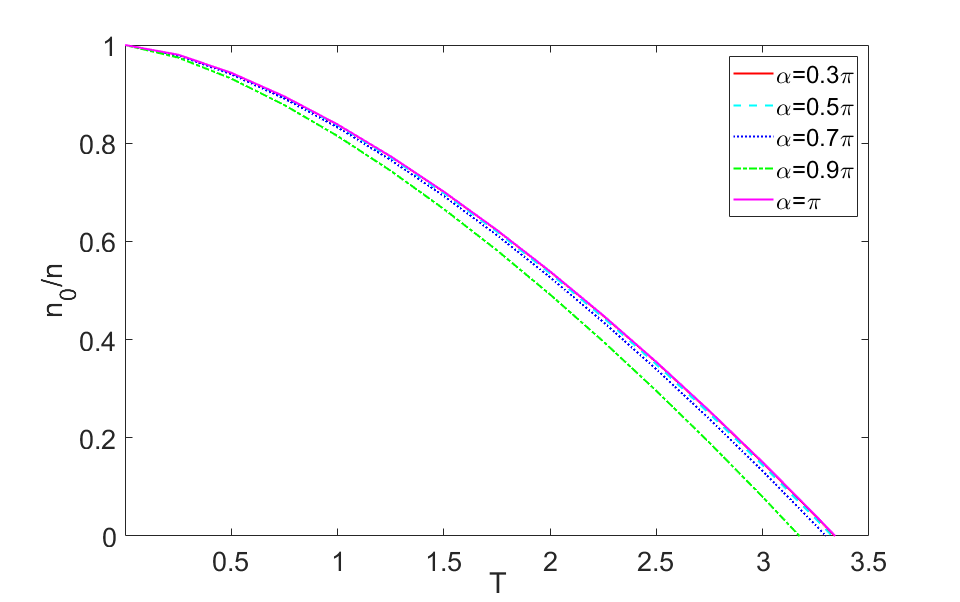}
\caption{\footnotesize (Color online) The ground state particle density $n_{0}/n$ versus $T$ with $L=100L_{0}$, with $\alpha=0.3\pi, 0.5\pi, 0.7\pi, 0.9\pi, \pi$ in the cubic box. Here $\frac{\hbar^2}{mk}n^{\frac{2}{3}}$ is taken as an unit of $T$.}\label{fig12}
\end{figure}

In order to study the influence of $\alpha$ on condensation under twisted boundary conditions, we numerically calculate $T_{c,L}$ and $n_{0}/n$ versus with $\alpha$ in the cases of $L=L_{0}$, $L=10L_{0}$, $L=100L_{0}$, respectively. And the numerical results are then plotted in Figs. ~\ref{fig7}, ~\ref{fig8}, ~\ref{fig9}, ~\ref{fig10}, ~\ref{fig11}, ~\ref{fig12} respectively.

As shown in Figs. ~\ref{fig7}, ~\ref{fig8} and ~\ref{fig9}, Obviously, in the interval $[0,2\pi]$, the change of $T_{c,L}$ with $\alpha$ is symmetric about $\alpha=\pi$, which can be derived from the Section 2. And in the interval $[0,\pi)$, $T_{c,L}$ gradually becomes lower, but when $\alpha$ reaches $\pi$, $T_{c,L}$ takes the maximum. This singular behavior is due to a significant fact that changes in boundary conditions disproportionately adds or removes single-states with the zero quantum numbers~\cite{tarasov2018anomalous}, and the contribution from that group of single-state to the non-condensed occupation is quite large. Compared with other boundary conditions, the antiperiodic boundary conditions remove more single-state with the zero quantum numbers, which can be derived in Section II. As a result, the fixed particle number density determines that the characteristic temperature $T_{c,L}$ takes its maximum at $\alpha=\pi$. Additionally, comparing Figs. ~\ref{fig7}, ~\ref{fig8} and ~\ref{fig9}, as $L$ increases until it is large enough, 
the influence of the boundary conditions on $T_{c,L}$ becomes negligible.

Apart from this, the condensate fraction $n_{0}/n$ versus $T$ with various $\alpha$ in different linear sizes are presented in Figs. ~\ref{fig10}, ~\ref{fig11}, ~\ref{fig12}. The condensate fraction become less sensitive to the boundary conditions as the $L$ increases. Particularly, Fig. ~\ref{fig12} demonstrates that the signature of BEC in finite systems become independent of the boundary conditions when the linear size is large enough. In the case of small volume systems, as $\alpha$ changes from 0 to $\pi$, the major difference is a retarded onset of the occupation of the ground state, while $\alpha=\pi$, antiperiodic boundary condition is more conductive for bosons to condensation.

As a consequence, in a system of small volume, the characteristic temperature and condensate fraction are sensitive to boundary conditions, especially in the case of antiperiodic boundary condition.

\section{Conclusions}\label{Section5}
In this paper, we have studied ideal Bose gases with fixed particle density confined in the cubic box. By means of the theoretical analyses, we derived the specific formulas of the Bose distribution confined in the cubic box. Through numerical calculation, we analyzed the influence of the finite-size and boundary conditions on characteristic temperature $T_{c,L}$ and condensate fraction $n_{0}/n$. We found that in the case of finite volume system, the smaller linear size can increase $T_{c,L}$ and $n_{0}$, the crossover behavior between the ground and excited states will be advanced as the linear size decreases. More importantly, the smaller the volume, the more sensitive $T_{c,L}$ and $n_{0}/n$ are to antiperiodic boundary conditions.

Additionally, superconductivity and superfluidity that have a lot in common with BEC can be described by similar theories. Therefore, the finite size effect is expected to have an important impact on superconductivity and superfluidity, which requires further research.




\section*{Acknowledgments}
This work is jointly supported by  the National Nature Science Foundation of China (Grants No. 12075117, No. 51721001, No. 11890702, No. 11625418, No. 11535005, No. 11690030), the National Major state Basic Research and Development of China (Grant No. 2016YFE0129300). Y.-L. W. was funded by the Natural Science Foundation of Shandong Province of China (Grant No. ZR2017MA010).

\normalem
\bibliographystyle{apsrev4-1}
\bibliography{boseeinstein}
\end{document}